\documentclass[review]{elsarticle}
\usepackage{lineno,hyperref}
\usepackage{color}
\usepackage[dvipsnames]{xcolor}
\usepackage[normalem]{ulem}
\pdfoutput=1
\modulolinenumbers[5]
\journal{Journal of \LaTeX\ Templates}

\begin{document}
\begin{frontmatter} 
\title{Electron pairing in model with two overlapping bands}
\tnotetext[mytitlenote]{Electron pairing in model with two overlapping bandss}
\author{Igor N. Karnaukhov}
\address{G.V. Kurdyumov Institute for Metal Physics, 36 Vernadsky Boulevard, 03142 Kiev, Ukraine}
\fntext[myfootnote]{karnaui@yahoo.com}

\begin{abstract}

We consider the two-band Hubbard model, where electrons from different bands interact through an on-site one- and two-particle hybridization.
 The proposed Hamiltonian makes it possible to construct an effective theory and answer the question of the nature of pairing: conduction electrons form pairs due to two-particle hybridization of electrons from different bands, compensating for the direct Hubbard repulsion between conduction electrons. 
 It is shown that an effective attraction between conduction electrons leads to $\eta$-pairing.  The electron-electron pairing mechanism explains the presence of superconductivity at high temperatures experimentally observed in hydrogen-rich materials at high pressure.
 \end{abstract}
 \begin{keyword}
\texttt  pairing \sep hybridization \sep high-temperature superconductivity
\end{keyword}
\end{frontmatter}
\section{Introduction}

The impressively high critical temperature, unconventional Cooper pairing bonding in cuprates and iron-based superconductors are topical issues in many-body physics. Research into high-temperature superconductivity \cite{1}  at high pressure has made it possible to achieve superconductivity at significant temperatures— 200 degrees Kelvin and higher \cite{2,3}.  A record of superconducting transition temperature $\sim 250 K$ has been achieved in {LaH}$_{10}$ \cite{4}. We are already talking about superconductors at room temperature.
 Nevertheless, we must acknowledge the fact that, despite such progress, we have not advanced in our understanding of the nature of high-temperature superconductivity, the physical mechanism of electron pairing, the formation of condensate of electron pairs.
It is obvious that such high temperatures of superconducting transition cannot be explained within the framework of electron-phonon mechanism of pairing (the Cooper pairing); most likely, the electron-electron pairing mechanism is realised. As in the case of the Cooper pairing, effective attraction between conduction electrons can only arise in the case of their indirect interaction through another subsystem, since direct interaction between electrons is always repulsive. It is desirable that such an electron-electron pairing mechanism be sufficiently simple to understand (such as  the Cooper pairing) and not require strict conditions.

Since cuprate superconductors are doped Mott insulators, it would be useful to propose a model that demonstrates both the properties of a Mott insulator and superconductivity.
In this work, we will attempt to propose  a mechanism of electron-electron pairing within the framework of a two-band electron model, each band in  which is described in the framework of the Hubbard model. Our study can be readily extended to band insulators.  It is clear that the on-site repulsion of electrons in the Hubbard model has nothing to do with the attraction of conduction electrons. We will show that two-particle hybridization between electrons of different bands can lead to the effective attraction between electrons of one band and the repulsion of electrons of the other band. 

High-temperature superconductors involve strongly hybridized states near the Fermi level.
From electronic band structure calculations of  hydrogen sulphide  H$_{3}$S follows that the hybridization between H and S states is strongly enhanced in a broader region around the Fermi energy,  
where  S$_{3p}$-H$_{1s}$  interaction dominates \cite{4}. In palladium hydride PdH the hybridization between H and Pd states is sufficiently large near the Fermi energy \cite{4}. The relevance of hybridization in the formation of the superconducting state was noted in \cite{IK1}.
Unlike the Cooper pairing, in which electrons form pairs with zero total momentum, electron pairing occurs at a momentum equal to $\pi$. This is called $\eta$-pairing \cite{Y}-\cite{6}. 
During $\eta$-pairing, electrons with momenta $\overrightarrow{k}$ and $\overrightarrow{\pi}- \overrightarrow{k}$ ($\overrightarrow{\pi}$ is vector) form pairs, which leads to the condensation of
electron pairs with $\pi$-momentum. This refers to spatially modulated superconductivity and charge density. In the Hubbard model with on-site repulsive interaction, $\eta$-pairing does not realized, so it is important to consider the interaction between electrons of different bands. 

 \section{The model Hamiltonian}

The Hamiltonian defines the two-band model, in which $s$- and $p$-electrons (interacting via the Hubbard interaction) have different nearest-neighbor hopping integrals, their energies are shifted relative to each other and $s$- and $p$- electrons interact through the on-site one- and two-particle hybridization. The Hamiltonian of the model has the following form  ${\cal H}={\cal H}_s+{\cal H}_p+{\cal H}_{s-p}$:
\begin{eqnarray}
&&{\cal H}_s=- \sum_{\sigma}\sum_{\textbf{j}, \textbf{1}} (c^\dagger_{\textbf{j}\sigma}c_{\textbf{j+1}\sigma} +c^\dagger_{\textbf{j+1}\sigma}c_{\textbf{j}\sigma})+U_s \sum_{j=1}^{N}  n^s_{\textbf{j}\uparrow}n^s_{\textbf{j}\downarrow},\nonumber \\&&
{\cal H}_p=- t\sum_{\sigma}\sum_{\textbf{j},\textbf{1}} (p^\dagger_{\textbf{j}\sigma}p_{\textbf{j+1}\sigma} +p^\dagger_{\textbf{j+1}\sigma}p_{\textbf{j}\sigma})+\varepsilon \sum_{j=1}^{N}  n^p_{\textbf{j}}+U_p \sum_{j=1}^{N}  n^p_{\textbf{j}\uparrow}n^p_{\textbf{j}\downarrow},\nonumber \\&&    
{\cal H}_{s-p}=v \sum_{\sigma}\sum_{{j}=1}^{N} (c^\dagger_{\textbf{j}\sigma}p_{\textbf{j}\sigma}+
p^\dagger_{\textbf{j}\sigma}c_{\textbf{j}\sigma})+
g \sum_{j=1}^{N} (c^\dagger_{\textbf{j}\uparrow}c^\dagger_{\textbf{j}\downarrow}p_{\textbf{j}\uparrow}p_{\textbf{j}\downarrow}+
p^\dagger_{\textbf{j}\downarrow}p^\dagger_{\textbf{j}\uparrow}c_{\textbf{j}\downarrow}c_{\textbf{j}\uparrow}),\nonumber \\
\label{eq:1}
\end{eqnarray}
where $c^\dagger_{\textbf{j}\sigma},c_{\textbf{j}\sigma}$ and  $p^\dagger_{\textbf{j}\sigma},p_{\textbf{j}\sigma}$ $(\sigma=\uparrow,\downarrow$) are the Fermi operators for $s$- and $p$-electrons defined on the site $\textbf{j}$, 
$1$ and $t\leq 1$ are the nearest-neighbor hopping integrals for $s$- and $p$-electrons and sums over the nearest lattice sites,
 $\varepsilon$ is an one-particle energy of $p$-electrons, which define the positions of $p$-subbands (valence and conduction) relative to $s$-conduction band,
$n^s_{\textbf{j}\sigma}=c^\dagger_{\textbf{j}\sigma}c_{\textbf{j}\sigma}$, $n^p_{\textbf{j}\sigma}=p^\dagger_{\textbf{j}\sigma}p_{\textbf{j}\sigma}$,  $n^{s,p}_{\textbf{j}}=n^{s,p}_{\textbf{j}\uparrow}+n^{s,p}_{\textbf{j}\downarrow}$ are the density operators for $s$- and $p$-electrons. $U_s$ and $U_p$ determine  the on-site Hubbard interactions within  bands. 
 Parameters $v$ and $g$  determine the magnitudes of the one- and two-particle on-site $s$-$p$ hybridization of electrons. $N$ is the total number of the lattice sites.

In the  single-impurity Anderson model \cite{7}, conduction electrons hybridize with localized electrons, which exist in three (one- and two-particle) states.  In the case, when the energy corresponding to states of an localized electron with different spin is outside the conduction band, conduction electrons do not hybridize with localized electrons in these states.  Conduction electrons hybridize with two localized electrons located at the same site, since the energy corresponding to this state lies within the conduction band. This two-particle hybridization is taken into account in the model Hamiltonian, which has been proposed in \cite{IK0}. 
The Hamiltonian (\ref{eq:1}) is generalization of the model \cite{IK0}, where band $s$-electrons interact with band $p$-electrons due to overlapping bands. The conduction $p$-band lies in the conduction $s$-band above the Fermi energy (see in Fig.1), $s$-electrons hybridize with $p$-electrons. Thus, unlike model \cite{IK0}, where $p$-band was considered as a flat band, in this case we are talking about the interaction between electrons from different bands (with different hopping integrals, narrow $p$- and wide $s$-conduction bands). The lower value of the Hubbard repulsion corresponds to $s$-electrons with a wide conduction band, i.e. $U_s<U_p$, while the value of $g$ most likely corresponds to an intermediate value, namely $U_s<g<U_p$.
In \cite{a1}, an extended version of the BCS model was proposed, taking into account that in more realistic compounds the electronic spectrum can be multi-band. A modification of the BCS theory was proposed from the perspective of two-band superconductivity. The proposed model (\ref{eq:1}) has nothing to do with the model \cite{a1}, as it is based on the interacting two-band Hubbard model, here a separate Hubbard band is not superconducting due to on-site repulsion of electrons. Below, we examine $\eta$-pairing within model (1) and show the feasibility of implementing a superconducting state. 

\begin{figure}[tp]
      \centering{\leavevmode}
\begin{minipage}[h]{.75\linewidth}
\center{
\includegraphics[width=\linewidth]{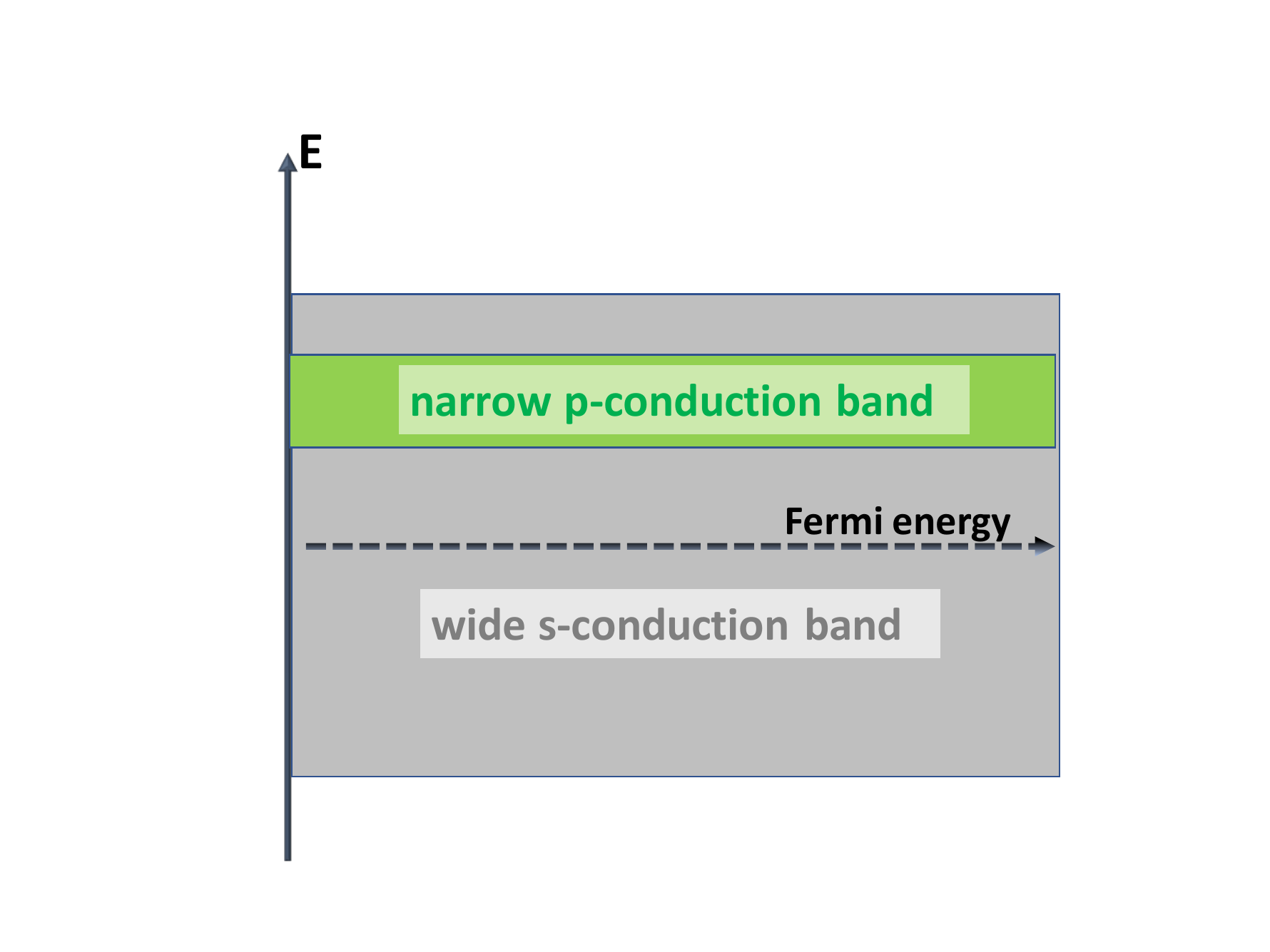} 
                  }
    \end{minipage}
\caption{The structure of the electron bands corresponding to the model is shown schematically.
} 
\label{fig:1}
\end{figure}

\begin{figure}[tp]
      \centering{\leavevmode}
\begin{minipage}[h]{.75\linewidth}
\center{
\includegraphics[width=\linewidth]{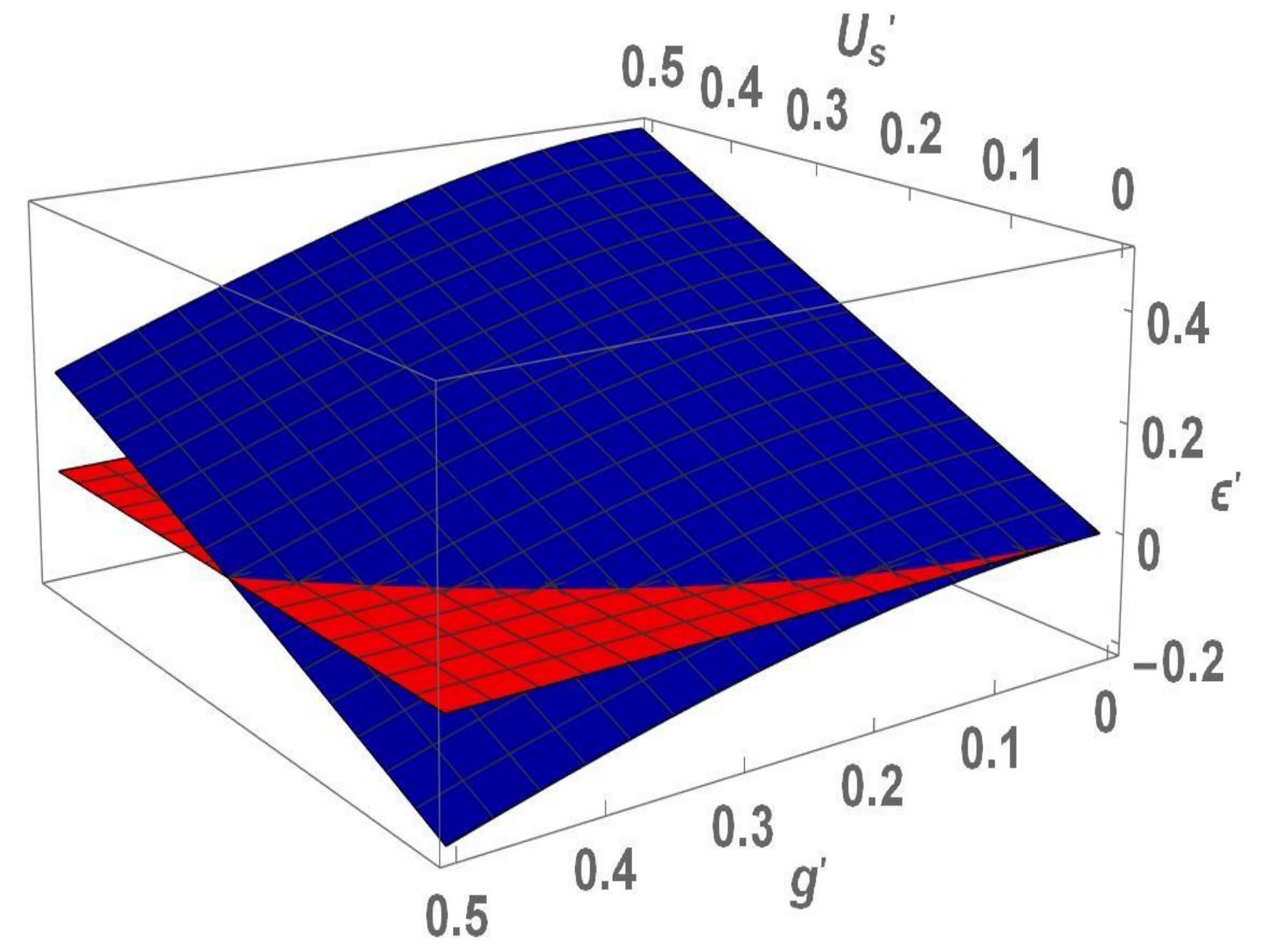} 
                  }
    \end{minipage}
\caption{Energy of electron pair as function of the on-site Hubbard and hybridization interactions in units of $\varepsilon'=2\varepsilon+U_p$:
$\epsilon'=\epsilon/\varepsilon'$, $U_s'=U_s/\varepsilon'$, $g'=g/\varepsilon'$.
} 
\label{fig:2}
\end{figure}

\section{Case of weak $s$-$p$ hybridisation}

Let's consider a simplified version of model (\ref{eq:1}), which does not take into account the on-site Hubbard interactions, redefine the model Hamiltonian  in the momentum presentation as ${\cal H}={\cal H}_{0}+{\cal H}_{hyb}$ 
\begin{eqnarray}
&&{\cal H}_0=\sum_\textbf{k} \epsilon_s (\textbf{k})n^s(\textbf{k})+
\sum_\textbf{k}[\epsilon_p (\textbf{k})+\varepsilon]n^p(\textbf{k}),
\nonumber \\&&
{\cal H}_{hyb}=v \sum_{\sigma}\sum_\textbf{k} [c^\dagger_{\sigma}(\textbf{k})p_{\sigma}(\textbf{k})+p^\dagger_{\sigma}(\textbf{k})c_{\sigma}(\textbf{k})]+
g  \sum_{\textbf{k}_1,\textbf{k}_2,\textbf{q}_1,\textbf{q}_2}\delta(\textbf{k}_1+\textbf{k}_2 -\textbf{q}_1-\textbf{q}_2) \nonumber \\&&
[c^\dagger_{\uparrow}(\textbf{k}_1) c^\dagger_{\downarrow}(\textbf{k}_2) p_{\uparrow}(\textbf{q}_1)p_{\downarrow}(\textbf{q}_2) +
p^\dagger_{\downarrow}(\textbf{k}_1)p^\dagger_{\uparrow}(\textbf{k}_2) c_{\downarrow}(\textbf{q}_1)c_{\uparrow}(\textbf{q}_2) ],
\label{eq:2}
\end{eqnarray}
where $\epsilon_s (\textbf{k})=-2(\cos k_x +\cos k_y +\cos k_z)$, $\epsilon_p (\textbf{k})=-2t(\cos k_x +\cos k_y +\cos k_z)$,
$\textbf{k}=\{k_x,k_y,k_z\}$ is a wave vector of an electron, 
$c^\dagger_{\textbf{j}\sigma}=\sum_{\textbf{k}}c^\dagger_{\sigma}(\textbf{k})\exp(i \textbf{kj})$ and $p^\dagger_{\textbf{j}\sigma}=\sum_{\textbf{k}}p^\dagger_{\sigma}(\textbf{k})\exp(i \textbf{kj})$, $n^p_{\sigma}(\textbf{k})=p^\dagger_{\sigma}(\textbf{k})p_{\sigma}(\textbf{k})$, $n^s_{\sigma}(\textbf{k})=c^\dagger_{\sigma}(\textbf{k})c_{\sigma}(\textbf{k})$, $n^{s,p}(\textbf{k})=n^{s,p}_{\uparrow}(\textbf{k})+n^{s,p}_{\downarrow}(\textbf{k})$.
The expressions for $\epsilon_{s,p}(\textbf{k)}$ are given for a simple cubic lattice, although the derivation of the effective Hamiltonian does not depend on the explicit form of the dispersion of $s$-and $p$-electrons.

Using the Schrieffer-Wolff transformation, the Hamiltonian  (\ref{eq:2}) is defined in the second order interaction in the tradition form ${\cal H}_{eff}={\cal H}_0+\frac{1}{2}[S,{\cal H}_{v}]$, where $S$-operator is equal to 
\begin{eqnarray}
&&
S=\sum_{\sigma}\sum_{\textbf{k}}\frac{v}{\epsilon_s(\textbf{k})-\epsilon_p(\textbf{k})-\varepsilon} [c^\dagger_{\sigma}(\textbf{k})p_{\sigma}(\textbf{k})- p^\dagger_{\sigma}(\textbf{k})c_{\sigma}(\textbf{k})]+\nonumber\\&&
\sum_{\textbf{k}_1,\textbf{k}_2,\textbf{q}_1,\textbf{q}_2}\frac {g}{\epsilon_s (\textbf{k}_1)+\epsilon_s (\textbf{k}_2)
-\epsilon_p (\textbf{q}_1)-\epsilon_p (\textbf{q}_2)-2\varepsilon} \delta(\textbf{k}_1+\textbf{k}_2 -\textbf{q}_1-\textbf{q}_2) \nonumber\\&&
[c^\dagger_{\uparrow}(\textbf{k}_1)c^\dagger_{\downarrow}(\textbf{k}_2)p_{\uparrow}(\textbf{q}_1) p_{\downarrow} (\textbf{q}_2) -
p^\dagger_{\downarrow}\textbf{(q}_1)p^\dagger_{\uparrow}(\textbf{q}_2)c_{\downarrow}(\textbf{k}_2)c_{\uparrow}(\textbf{k}_1)].
\label{eq:3}
\end{eqnarray}

The effective Hamiltonian ${\cal H}_{eff}={\cal H}_0+{\cal H}_{v}+{\cal H}_{g}+{\cal H}_{v-g}$ includes four terms, here ${\cal H}_{v}$ is a one-particle term that shifts the energy of the electron liquid, defined by one-particle hybridization:
\begin{eqnarray}
&&
{\cal H}_{v}=\sum_{\textbf{k}}\frac{v^2}{\epsilon_s(\textbf{k})-\epsilon_p(\textbf{k})-\varepsilon} [n^s(\textbf{k})-n^p(\textbf{k})].
\label{eq:4}
\end{eqnarray}
Next term ${\cal H}_{g}$ quadratic on $g$ define two-particle interaction of $s$- (the term ${\cal H}^s_{g}$) and $p$- (the term ${\cal H}^p_{g}$) electrons, here ${\cal H}_{g}={\cal H}^s_{g}+{\cal H}^p_{g}$:

\begin{eqnarray}
&&
{\cal H}^s_{g}=\frac{g^2}{2}\sum_{\textbf{k}_1,\textbf{k}_1',\textbf{k}_2,\textbf{k}_2', \textbf{q}_1,\textbf{q}_1',\textbf{q}_2,\textbf{q}_2'}
\frac{\delta(\textbf{k}_1+\textbf{k}_2 -\textbf{q}_1-\textbf{q}_2) \delta(\textbf{k}_1'+\textbf{k}_2' -\textbf{q}_1'-\textbf{q}_2') }
{\epsilon_s (\textbf{k}_1)+\epsilon_s (\textbf{k}_2)-\epsilon_p (\textbf{q}_1)-\epsilon_p (\textbf{q}_2)-2\epsilon} \nonumber\\&& 
\{c^\dagger_{\uparrow}(\textbf{k}_1)c_{\uparrow}(\textbf{k}_1')c^\dagger_{\downarrow}(\textbf{k}_2)c_{\downarrow}(\textbf{k}_2') [p_{\uparrow}(\textbf{q}_1)p^\dagger_{\uparrow}(\textbf{q}_1') \delta (\textbf{q}_2-\textbf{q}_2')-
p^\dagger_{\downarrow}(\textbf{q}_2')p_{\downarrow}(\textbf{q}_2) \delta (\textbf{q}_1-\textbf{q}_1')]+
\nonumber\\&& 
c^\dagger_{\uparrow}(\textbf{k}_1')c_{\uparrow}(\textbf{k}_1)c^\dagger_{\downarrow}(\textbf{k}_2')c_{\downarrow}(\textbf{k}_2) [p_{\uparrow}(\textbf{q}_1')p^\dagger_{\uparrow}(\textbf{q}_1) \delta (\textbf{q}_2-\textbf{q}_2')-
p^\dagger_{\downarrow}(\textbf{q}_2)p_{\downarrow}(\textbf{q}_2') \delta (\textbf{q}_1-\textbf{q}_1')]
\}.\nonumber\\
\label{eq:5}
\end{eqnarray}

Introducing $ \rho^{s,p}(\textbf{k})=<n^{s,p}(\textbf{k})>$  the term (\ref{eq:5}) is simplified as
\begin{eqnarray}
&&
{\cal H}^s_{g}=\frac{g^2}{2}\sum_{\textbf{k}_1,\textbf{k}_1',\textbf{k}_2,\textbf{k}_2',\textbf{q}_1,\textbf{q}_2}
\frac{\delta(\textbf{k}_1+\textbf{k}_2 -\textbf{q}_1-\textbf{q}_2) \delta(\textbf{k}_1'+\textbf{k}_2'-\textbf{q}_1-\textbf{q}_2 )}{\epsilon_s (\textbf{k}_1)+\epsilon_s (\textbf{k}_2)-\epsilon_p (\textbf{q}_1)-\epsilon_p (\textbf{q}_2)-2\epsilon} \nonumber\\&& 
[1-\rho^p(\textbf{q}_1)][c^\dagger_{\uparrow}(\textbf{k}_1)c_{\uparrow}(\textbf{k}_1')c^\dagger_{\downarrow}(\textbf{k}_2)c_{\downarrow} (\textbf{k}_2')+ c^\dagger_{\uparrow}(\textbf{k}_1')c_{\uparrow}(\textbf{k}_1)c^\dagger_{\downarrow}\textbf{(k}_2') c_{\downarrow}(\textbf{k}_2)].
\label{eq:6}
\end{eqnarray}

The term ${\cal H}^s_{g}$ determines the two-particle scattering of $s$-electrons, which is the result of the two-particle  $s$-$p$ hybridization. The term  ${\cal H}^p_{g}$ is determined by formulas (\ref{eq:5}) and (\ref{eq:6}) with the replacement of the indices defining the type of electrons from $s\to p$ ($c \to p$ for operators) and the energies $\epsilon_s(\textbf{k})\to \epsilon_p (\textbf{k})+\varepsilon$. Term  ${\cal H}_{v-g}$ has a structure $v g (c^\dagger c^\dagger c  p +p^\dagger p^\dagger p  c...)$, it determines the one-particle hybridization in  two-electron state, its contributes to ${\cal H}_{eff}$ of  a higher order in interaction than that of the ones mentioned above.

Unlike one-particle hybridization, which does not lead to the emergence of interaction between electrons, two-particle hybridization leads to the emergence of electron interaction. According to the calculations obtained (\ref{eq:5}),(\ref{eq:6}), the effective interaction between $s$-electrons is attractive when the $p$-band is located above the Fermi level, at $\varepsilon >0$ (this band structure configuration is shown in Fig.1). The magnitude of attractive interaction is maximal at 
$\epsilon_s(\textbf{k}_1)=-\epsilon_s (\textbf{k}_2)$, when electrons form $\eta$-pairing \cite{Y,IK0}. Below, the mechanism of the $\eta$ pairing is considered within the framework of model (\ref{eq:1}), without taking into account the one-particle hybridization between $s$- and $p$-electrons, since, as shown above, one-particle hybridization does not lead to effective interaction between electrons.

\section{$\eta$-pairing for the model  with $v=0$}

In the model (\ref{eq:1}) $\eta$-pairing  is determined by two pairing operators $\eta=\sum_\textbf{j}\eta_\textbf{j}$, 
 $\eta_\textbf{j}=\exp (i \overrightarrow{\pi} \overrightarrow{j})c^\dagger_{\textbf{j}\uparrow}c^\dagger_{\textbf{j}\downarrow}$  and 
$\mu=\sum_\textbf{j} \mu_{\textbf{j}}$, $\mu_{\textbf{j}}=\exp (i \overrightarrow{\pi} \overrightarrow{j})p^\dagger_{\textbf{j}\uparrow} p^\dagger_{\textbf{j}\downarrow}$ for $s$- and $p$-electrons \cite{IK2}. 
These operators describe $s$- and $p$-electron pairs, as these states are bounded  together through interaction. The pairing operators satisfy the following relations
\begin{eqnarray}
&&[{\cal H},\eta]=U_s\eta-g\mu+g G_{\mu},\nonumber \\ 
&&[{\cal H},\mu]=(2\varepsilon +U_p)\mu -g\eta+ g G_{\eta},
\label{eq:7}
\end{eqnarray}
where
$ G_{\mu}=\sum_\textbf{j} \mu_{\textbf{j}} n^s_{\textbf{j}}$, $G_{\eta}=\sum_\textbf{j} \eta_{\textbf{j}} n^p_{\textbf{j}}$, 

Coordinate-symmetric operators $\eta_\textbf{j}$ and $\mu_\textbf{j}$ correspond to non-trivial solutions of the pairing operators. $\eta_\textbf{j}n^p_\textbf{j}$ and $\mu_\textbf{j}n^c_\emph{j}$ are also coordinate-symmetric, which also implies the symmetry of electron density operators: $n^s_\textbf{j}=n^s_{-\textbf{j}}$ and $n^p_\textbf{j}=n^p_{-\textbf{j}}$. Coordinate-antisymmetric $\eta_\textbf{j}$, $\mu_\textbf{j}$,  $\eta_\textbf{j}n^p_\textbf{j}$ and $\mu_\textbf{j}n^s_\emph{j}$ operators correspond to zero solutions for the pairing operators.

Higher-order equations for electron operators have the following form:
\begin{eqnarray}
&&[{\cal H},G_\eta]=U_s G_{\eta}+2 g Q_{\mu}-\delta G_\eta,
\nonumber \\ 
&&[{\cal H},G_\mu]=(2\varepsilon+U_p) G_\mu + 2 g Q_{\eta}-\delta G_\mu,
\label{eq:8}
\end{eqnarray}
where 
$Q_{\mu}=\sum_{\textbf{j}}\mu_{\textbf{j}}n^s_{\textbf{j}\uparrow}
n^s_{\textbf{j}\downarrow}$, $Q_{\eta}=\sum_{\textbf{j}} \eta_{\textbf{j}}n^p_{\textbf{j}\uparrow}
n^p_{\textbf{j}\downarrow}$,

$\delta G_\eta =t\sum_{\sigma} \sum_{\textbf{j,1}}(p^\dagger_{\textbf{j+1}\sigma}p_{\textbf{j}\sigma}-p^\dagger_{\textbf{j} \sigma}p_{\textbf{j+1}\sigma}) \eta_{\textbf{j}}+\sum_{\textbf{j,1}} \exp (i \overrightarrow{\pi} \overrightarrow{j}) (c^\dagger_{\textbf{j+1}\uparrow}c^\dagger_{\textbf{j}\downarrow}+ c^\dagger_{\textbf{j}\uparrow}c^\dagger_{\textbf{j+1}\downarrow})
 n^p_{\textbf{j}} $;
   
$\delta G_\mu =\sum_{\sigma}\sum_{\textbf{j,1}} (c^\dagger_{\textbf{j+1}\sigma}c_{\textbf{j}\sigma}-
   c^\dagger_{\textbf{j}\sigma}c_{\textbf{j+1}\sigma}) \mu_{\textbf{j}}+
 t\sum_{\textbf{j,1}}\exp (i \overrightarrow{\pi} \overrightarrow{j}) (p^\dagger_{\textbf{j+1}\uparrow}p^\dagger_{\textbf{j}\downarrow}+
 p^\dagger_{\textbf{j}\uparrow}p^\dagger_{\textbf{j+1}\downarrow}) n^s_{\textbf{j}} $.

and 
\begin{eqnarray}
&&[{\cal H},Q_\eta]= U_s Q_{\eta}+g Q_{\mu}-\delta Q_{\mu}, \nonumber \\ 
&&[{\cal H},Q_\mu]=(2\varepsilon+U_p)  Q_{\mu}+g Q_{\eta}-\delta Q_{\eta},
\label{eq:9}
\end{eqnarray}
where 
$\delta Q_{\mu} =t\sum_{\textbf{j,1}}[
(p^\dagger_{\textbf{j+1}\uparrow}p_{\textbf{j}\uparrow}-p^\dagger_{\textbf{j}\uparrow}p_{\textbf{j+1}\uparrow})
n^p_{\textbf{j}\downarrow}+(p^\dagger_{\textbf{j+1}\downarrow}p_{\textbf{j}\downarrow}- p^\dagger_{\textbf{j}\downarrow}p_{\textbf{j+1}\downarrow})n^p_{\textbf{j}\uparrow}]
\eta_{\textbf{j}}+
\sum_{\textbf{j,1}}\exp(i\overrightarrow{\pi}\overrightarrow{j})(c^\dagger_{\textbf{j+1}\uparrow}c^\dagger_{\textbf{j}\downarrow}+ c^\dagger_{\textbf{j}\uparrow}c^\dagger_{\textbf{j+1}\downarrow})n^p_{\textbf{j}\uparrow}n^p_{\textbf{j}\downarrow}$,

$\delta Q_{\eta} =\sum_{\textbf{j,1}}[
(c^\dagger_{\textbf{j+1}\uparrow}c_{\textbf{j}\uparrow}-c^\dagger_{\textbf{j}\uparrow}c_{\textbf{j+1}\uparrow})
n^s_{\textbf{j}\downarrow}+(c^\dagger_{\textbf{j+1}\downarrow}c_{\textbf{j}\downarrow}- c^\dagger_{\textbf{j}\downarrow}c_{\textbf{j+1}\downarrow})n^s_{\textbf{j}\uparrow}]
\mu_{\textbf{j}}+ 
t\sum_{\textbf{j,1}}\exp(i\overrightarrow{\pi} \overrightarrow{j})(p^\dagger_{\textbf{j+1}\uparrow}p^\dagger_{\textbf{j}\downarrow}+ p^\dagger_{\textbf{j}\uparrow}p^\dagger_{\textbf{j+1}\downarrow})n^s_{\textbf{j}\uparrow}n^s_{\textbf{j}\downarrow}$.

In the Methods,
 we  showed that the summands $\delta G_{\eta}, \delta G_{\mu}, \delta Q_{\eta}, \delta Q_{\mu}$ in  Eqs (\ref{eq:8}),(\ref{eq:9})  are equal to zero in the case when the local operators  $\eta_{\textbf{j}},\mu_{\textbf{j}},n^s_{\textbf{j}\sigma}, n^p_{\textbf{j}\sigma}, n^s_{\textbf{j}\sigma}\mu_{\textbf{j}}$, 
$n^p_{\textbf{j}\sigma}\eta_{\textbf{j}}, n^s_{\textbf{j}\uparrow} n^s_{\textbf{j}\downarrow}, n^p_{\textbf{j}\uparrow}n^p_{\textbf{j}\downarrow}$ 
are coordinate-symmetric. The operators $\eta,\mu, G_\eta,G_\mu,Q_\eta, Q_\mu$ are defined by the sum over lattice sites of coordinate-symmetric local operators, therefore  
there are trivial solutions equal to zero for the coordinate-antisymmetric parts of the above local operators. We are only dealing with the coordinate-symmetric operators $\eta_\textbf{j}$, $\mu_{\textbf{j}}$, $n^s_{\textbf{j}\sigma}$ and $n^p_{\textbf{j}\sigma}$.

Taking into account that the terms $\delta G_\eta, \delta G_\mu, \delta Q_\eta, \delta Q_\mu$ are equal to zero, the following solutions for energy of electron pairs follow from Eqs (\ref{eq:7})-(\ref{eq:9}) $\epsilon_\pm =\frac {1}{2}[2\varepsilon+U_p+U_s \pm \sqrt{4g^2+(2\varepsilon+U_p-U_s)^2}]$.
In the case of weak interaction for $g<< 2\varepsilon+U_p-U_s$  and $2\varepsilon+U_p>U_s$ the energies are equal $\epsilon_+ =2\varepsilon+U_p+ \frac{g^2}{2\varepsilon+U_p-U_s}$,  $ \epsilon_-=U_s-\frac{g^2}{2\varepsilon+U_p-U_s}$ and for $U_s=2\varepsilon+U_p$  $\epsilon_\pm =U_s\pm g$, for $g>\sqrt{U_s (2\varepsilon+U_p)}$ $\epsilon_-<0$  (see in Fig. 2). In the case $\epsilon_-<0$  s-electrons form pairs (or $\eta$-pairing) forming a superconducting state. 
It should be emphasized that the energy of electron pair does not depend on the $s-,p-$band widths, it is determined by the interaction constants (in particular, the $p$-band can also be a flat band). 
In the case when the $p$-bands are flat \cite{IK2}, the electron spectrum is shown in Fig.1, where a high-energy $p$-flat band lies within  $s$-conduction band, and  a low-energy $p$-flat band is located below the $s$-conduction band; thus, the condition $v=0$ is realized.
As noted above, the condition $g>U_s$ can be achieved, therefore the considered pairing mechanism can be realized under real conditions. In the presence of interaction, electron pairing occurs, which leads to a transition to a superconducting phase that is distinct from the BCS state. These results may obviously be generalized to more complex band structures (in particular, band insulators).
It is possible that in experiments on achieving superconductivity at high pressure, these very necessary conditions are realized, namely an increase in the hybridization of electronic states with increasing pressure \cite{4}.

\section {Methods}

Below, we will examine in detail the solutions to Eqs (\ref{eq:7})-(\ref{eq:9}) for pairing operators. 
 The pairing operators determine the singlet pairing of electrons in real space, so the functions corresponding to the operators in Eqs (\ref{eq:7})-(\ref{eq:9})  are coordinate-symmetric or spin-antisymmetric. We consider solutions of the equations based on singlet functions or their corresponding operators. Let's use the Fourier representation for operators:\\
$\eta_{\textbf{j}}=\sum _{\textbf{k}}\eta(\textbf{k})\exp (i\textbf{kj})$;
$\mu_{\textbf{j}}=\sum _{\textbf{k}}\mu(\textbf{k})\exp (i\textbf{kj})$;\\
$n^s_{\textbf{j}}=\sum _{\textbf{k}}n^s(\textbf{k})\exp (i\textbf{kj})$;
$n^p_{\textbf{j}}=\sum _{\textbf{k}}n^p(\textbf{k})\exp (i\textbf{kj})$;\\
 $\eta_{\textbf{j}}=\sum _{\textbf{k,k'}}c^\dagger_{\uparrow} (\textbf{k})c^\dagger_{\downarrow} (\textbf{-k'})\exp [i(\textbf{k-k'+\overrightarrow{\pi})j}]=
 \sum _{\textbf{k,k'}}\eta({\textbf{k,-k'}})\exp [i\textbf{(k-k'+\overrightarrow{\pi})j}]$;\\
$\mu_{\textbf{j}}=\sum _{\textbf{k,k'}}p^\dagger_{\uparrow} (\textbf{k})p^\dagger_{\downarrow} (\textbf{-k'})\exp [i(\textbf{k-k'+\overrightarrow{\pi})j}]=
 \sum _{\textbf{k,k'}}\mu({\textbf{k,-k'}})\exp [i\textbf{(k-k'+\overrightarrow{\pi})j}]$;\\
 $n^{s}_{\textbf{\textbf{j}}\sigma}=\sum _{\textbf{k,k'}} c^\dagger_{\sigma}({\textbf{k}})c_{\sigma}({\textbf{k'}})
 \exp [i\textbf{(k-k')j}]=\sum _{\textbf{k,k'}} n^{s}({\textbf{k,k'}})\exp [i\textbf{k-k')j}]$;\\
  $n^{p}_{\textbf{\textbf{j}}\sigma}=\sum _{\textbf{k,k'}}
 p^\dagger_{\sigma}({\textbf{k}})p_{\sigma}({\textbf{k'}})
 \exp [i\textbf{(k-k')j}]=\sum _{\textbf{k,k'}}n^{p}({\textbf{k,k'}})\exp [i\textbf{(k-k')j}]$.
  
First of all, we consider two terms of the operator $\delta G_{\eta}$: $\delta G_{\eta}=t S^{(1)}_{\eta}+S^{(2)}_{\eta}=\sum_{\textbf{j}}[t S^{(1)}_{\eta}(\textbf{j})+S^{(2)}_{\eta}(\textbf{j})]$, where

  $S^{(1)}_{\eta}(\textbf{j})=\sum_{\sigma}\sum_{\textbf{1}}(p^\dagger_{\textbf{j+1}\sigma}p_{\textbf{j}\sigma}-p^\dagger_{\textbf{j} \sigma}p_{\textbf{j+1}\sigma})\eta_{\textbf{j}} $ and 
  
  $S^{(2)}_{\eta}(\textbf{j})=\sum_{\textbf{j,1}} \exp (i \overrightarrow{\pi j}) (c^\dagger_{\textbf{j+1}\uparrow}c^\dagger_{\textbf{j}\downarrow}+ c^\dagger_{\textbf{j}\uparrow}c^\dagger_{\textbf{j+1}\downarrow})
 n^p_{\textbf{j}}$.
 
Below we calculate the coordinate-symmetric parts of the operators $S^{(1)}_{\eta}(\textbf{j})$ and $S^{(2)}_{\eta}(\textbf{j})$ (their coordinate-antisymmetric terms do not contribute to $\delta G_{\eta}$), the corresponding equations take the following form:

$S^{(1)}_{\eta}=\sum_{\textbf{k,k',1}}[\exp(i\textbf{k1})p^\dagger_{\sigma}(\textbf{k})p_{\sigma}(\textbf{k'})- \exp(-i\textbf{k'1}) p^\dagger_{\sigma} (\textbf{k})p_{\sigma}(\textbf{k'})]  \eta(\textbf{k'-k})=2\sum_{\textbf{k,k'}}[\cos (\textbf{k1})- \cos (\textbf{k'1}) ]  p^\dagger_{\sigma}(\textbf{k})p_{\sigma}(\textbf{k'}) \eta(\textbf{k'-k})$. For the coordinate-symmetric  $\eta(\textbf{k-k'})= \eta(\textbf{k'-k})$ and $n^p(\textbf{k,k'})=n^p(\textbf{k',k})$ (the relations  follow from coordinate-symmetry $\eta_\textbf{j}$ and $n^p_{\textbf{j}\sigma}$)  we obtain that the coordinate-symmetric part of the operator $S^{(1)}_{\eta}$ is equal to zero ($\eta$-pairing corresponds to coordinate-symmetric operators).

$S^{(2)}_{\eta}=\frac{1}{2}\sum_{\textbf{j1}} (c^\dagger_{\textbf{j+1}\uparrow}c^\dagger_{\textbf{j}\downarrow}+ c^\dagger_{\textbf{j}\uparrow}c^\dagger_{\textbf{j+1}\downarrow}) \exp (i \overrightarrow{\pi j}) n^p_{\textbf{j}} =
\\
\sum_{\textbf{k,k'}}  
[\sin(\textbf{k1})-\sin(\textbf{k'1})]c^\dagger_{\uparrow}(\textbf{k}-\frac{1}{2}\pi) c^\dagger_{\downarrow}(-\textbf{k'}-\frac{1}{2}\pi)
n^p_{\textbf{k'-k}}$.
 According to the symmetry   $\eta (\textbf{k}-\frac{1}{2}\overrightarrow{\pi}, -\textbf{k'}-\frac{1}{2}\overrightarrow{\pi})=\eta (\textbf{k'}-\frac{1}{2}\overrightarrow{\pi},- \textbf{k}-\frac{1}{2}\overrightarrow{\pi})$ and $n^p(\textbf{k'-k})= n^p(\textbf{k-k'})$   we obtain that the coordinate-symmetric part of the operator $S^{(2)}_{\eta}$ is equal to zero.
 
Similarly to operator $\delta G_{\eta}$, we obtain the same result for operator $\delta G_{\mu}$: coordinate-symmetric parts of operator $\delta G_{\mu}$ is equal to zero due to following symmetry relations 
$\mu (\textbf{k-k'})=\mu (\textbf{k'-k})$, $n^s (\textbf{k,k'})=n^s (\textbf{k',k})$, $\mu (\textbf{k}-\frac{1}{2}\overrightarrow{\pi}, -\textbf{k'}-\frac{1}{2}\overrightarrow{\pi})=\mu (\textbf{k'}-\frac{1}{2}\overrightarrow{\pi},- \textbf{k}-\frac{1}{2}\overrightarrow{\pi})$ and $n^s(\textbf{k'-k})= n^s(\textbf{k-k'})$.  The structure of expressions for operators $\delta Q_{\eta}$ and $\delta Q_{\mu}$  is similar to that for operators $\delta G_{\eta}$ and $\delta G_{\mu}$, the coordinate-symmetric parts of operators $\delta Q_{\eta}$ and $\delta Q_{\mu}$ are equal to zero for coordinate-symmetric  operators $n^p_{\textbf{j}\sigma}$ and $n^p_{\textbf{j}\sigma}\eta_{\textbf{j}}$, $\eta_{\textbf{j}}$ and $n^p_{\textbf{j}\uparrow}n^p_{\textbf{j}\downarrow}$ (for $\delta Q_{\eta}$), $n^s_{\textbf{j}\sigma}$ and $n^s_{\textbf{j}\sigma}\mu_{\textbf{j}}$, $\mu_{\textbf{j}}$ and $n^s_{\textbf{j}\uparrow}n^s_{\textbf{j}\downarrow}$ (for $\delta Q_{\mu}$).
  
\section{Conclusion} 
 
To demonstrate the nature of the new electron-electron pairing mechanism, which, as far as is known, has not been documented before, a fairly simple two-band model was considered, in which the attraction between conduction electrons in the singlet channel is due to two-particle hybridization between electrons of different bands. 
The proposed pairing mechanism underlying superconductivity differs from traditional Cooper pairing in that it involves
electron pairing, which is realised as $\eta$-pairing.  Although $p$-electrons may not directly participate in the phenomenon of superconductivity,  $s$-electrons can benefit from their interaction with $p$-electrons.
The advantage of the mechanism of electron pairing is obvious, as it allows one to explain the abnormally high critical temperatures achieved in compounds under high pressure. The fact that $\eta$-pairing occurs in this case can be one of the criteria for experimentally testing the implementation of this unconventional pairing mechanism in superconductors.  This study can serve as a guide for explaining the nature of high-temperature superconductivity.

\subsection*{Author contributions statement}
I.K. is the author of the manuscript
\subsection*{Additional information}
The author declares no competing financial interests. 
\subsection*{Availability of Data and Materials}
All data generated or analysed during this study are included in this published article.


\begin{thebibliography}{31}
\bibitem{1} J.G.Bednorz and  K.A.M\"{u}ller, Possible high-$T_c$ superconductivity in the Ba-La-Cu-O system, Z.Phys.B \textbf{64} (1986) 189
\bibitem{2}     
 J.A.Flores-Livas, L.Boeri, A.Sanna et al. A perspective on conventional high-temperature superconductors at high pressure: Methods and materials, Phys. Rep.  \textbf{856} (2020) 1–78
\bibitem{3}         
P.Kong , V.S.Minkov, M.A.Kuzovnikov et al. Superconductivity up to $243$ K in the yttrium-hydrogen system under high pressure, Nat. Commun. \textbf{12} (2021) 5075;
I.A.Troyan, D.V.Semenok, A.G.Kvashnin  et al. Anomalous High-Temperature Superconductivity in $YH_6$,  Adv. Mater \textbf{ 33} (2021) 2006832 
\bibitem{4} 
 J.A.Flores-Livas, L.Boeri, A.Sanna,  G.Profeta, R.Arita, M.Eremets, A perspective on conventional high-temperature superconductors at high pressure: Methods and materials, Phys. Rep. \textbf{ 856 }, (2020) 1–78
 \bibitem{IK1}
I.N.Karnaukhov, Hybridized mechanism of pairing and the heavy fermion state: Exactly solvable two-band model with strong hybridized interactions, Phys. Rev. B \textbf{72}
 (2005) 092503 
\bibitem{Y} C.N.Yang, $\eta$-pairing and off-diagonal long-range order in a Hubbard model. Phys.Rev.Lett. \textbf{63}, (1989) 2144
    \bibitem{5} 
A.Lowe, M.Ortuno, I.V.Yurkevich, Topological phase transition in superconductors with mirror symmetry,
Journal of Physics: Condensed Matter, \textbf{32} (2019) 035603
\bibitem{6} 
H.Betsuyaku, $\eta$ pairing and superconductivity in the negative-U Hubbard model, Phys.Rev.B, \textbf{44}, (1991) 871
\bibitem{7} 
P.W.Anderson, Localized Magnetic States in Metalsm,  Phys.Rev. \textbf{124}, (1961) 41
\bibitem{IK0}
 I.N.Karnaukhov, {Solution of the Anderson chain with two-particle hybridization of localized and itinerant electrons},	arXiv:2509.20188 [cond-mat.str-el] (2025)
 \bibitem{a1}
 H.Suhl, B.T.Matthias, and L.R.Walker, Bardeen-Cooper-Schrieffer theory of superconductivity in the case of overlapping bands,
 Phys.Rev.Lett. \textbf{3} (1959) 552
 \bibitem{IK2}
 I.N.Karnaukhov, {$\eta$-pairing in the model with two-particle hybridization of conduction and localized electrons},		arXiv:2510.12349 [cond-mat.str-el] (2025)
 
 \end{thebibliography}
\end{document}